\documentclass[12pt,preprint]{aastex}

\shorttitle{An Intermediate Burst from SGR1900+14 with FREGATE}
\shortauthors{Olive et al.}

\newcommand\sgr{SGR1900+14}

\newcommand\msol{$\sf M_\odot$}

\newcommand\chisq{$\chi^2$}

\begin{document}

\title{Time-resolved X-ray spectral modeling of an intermediate burst from \sgr~observed by HETE-2/FREGATE
and WXM}

\author{J.-F. Olive}
\affil{Centre d'Etude Spatiale des Rayonnements, 
       9, avenue du Colonel Roche, 31029 Toulouse Cedex 4, France}
\email{Jean-Francois.Olive@cesr.fr}

\author{K. Hurley}
\affil{University of California at Berkeley,
       Space Sciences Laboratory,
       7 Gauss Way,
       Berkeley, CA 94720-7450}

\author{Takanori Sakamoto}
\affil{Tokyo Institute of Technology,
  2-12-1 Ookayama, Meguro-ku, Tokyo 152-8551, Japan}

\author{J.-L. Atteia}
\affil{Laboratoire d'Astrophysique,
        Observatoire Midi-Pyr\'en\'ees,
       14, avenue E. Belin,
       31400 Toulouse,France}

\author{G. Crew, G. Ricker}
\affil{Center for Space Research, Massachusetts Institute of Technology, 70 Vassar Street,
Cambridge, MA, 02139}

\author{G. Pizzichini}
\affil{Consiglio Nazionale delle Ricerche, IASF, Sezione di Bologna, via Piero
Gobetti 101, 40129 Bologna, Italy}

\author{C. Barraud}
\affil{Laboratoire d'Astrophysique,
        Observatoire Midi-Pyr\'en\'ees,
       14, avenue E. Belin,
       31400 Toulouse,France}

\author{Nobuyuki Kawai}
\affil{Tokyo Institute of Technology,
  2-12-1 Ookayama, Meguro-ku, Tokyo 152-8551, Japan, and \\
  RIKEN (The Institute of Physical and Chemical Research),
  2-1 Hirosawa, Wako, Saitama 351-0198, Japan}


\begin{abstract}

We present a detailed analysis of a 3.5 s long burst from SGR1900+14 which
occurred on 2001 July 2.  The 2-150 keV time-integrated energy spectrum is well described by the sum
of two blackbodies whose temperatures are approximately 4.3 and 9.8 keV.  The time-resolved
energy spectra are similarly well fit by the sum of two blackbodies.
The higher temperature blackbody evolves with time in a manner consistent 
with a shrinking emitting surface. The interpretation of these results
in the context of the magnetar model suggests that the two blackbody fit
is an approximation of an absorbed, multi-temperature spectrum expected on theoretical grounds
rather than a physical description of the emission. 
If this is indeed the case, our data provide further evidence for a strong
magnetic field, and indicate that the entire neutron
was radiating during most of the burst duration.

\end{abstract}



\keywords{stars: individual (\sgr) - stars: neutron}

\section{Introduction}
\label{intro}


The soft gamma repeater \sgr\ was discovered in 1979 when it emitted 3
short bursts of soft gamma-rays in 3 days (Mazets et al. 1979).  Its
next recorded appearance occurred some 13 years later (Kouveliotou et
al. 1993).  Attempts were made over the years to obtain a precise
position for the source and identify its counterpart (Hurley et
al. 1994; Vasisht et al.  1994; Hurley et al. 1996), but this remained
elusive until 1998, when the source entered a new period of activity,
allowing it to be localized accurately by the interplanetary network
(IPN: Hurley et al. 1999a).  The precise source location was found to
be consistent with that of a previously identified ROSAT quiescent
X-ray source (Vasisht et al. 1994; Hurley et al. 1996).  Observations
with ASCA further revealed that the source had a 5.16 s period (Hurley
et al. 1999b), and observations with RXTE demonstrated that the period
was increasing rapidly ($\rm 1.1 \times 10^{-10}$ s s$^{-1}$,
Kouveliotou et al. 1999). The counterpart to \sgr~has not been found
yet.  If it is associated with the Galactic supernova remnant
G42.8+0.6, it could be as close as 5 kpc (Hurley et al. 1999b).  However, the source position 
also appears to be very close to a
cluster of high mass stars, and it has been proposed that this may be
the birthplace of the neutron star (Vrba et al. 2000).  If so, its
distance could be roughly 12-15 kpc. In all this paper we adopt a distance
of 10 kpc. 

The 1998 activity of \sgr~culminated in the giant flare of August 27
(Frail et al. 1999; Hurley et al. 1999c; Feroci et al. 1999; Mazets et
al. 1999), which was followed by numerous, smaller bursts
(e.g. Ibrahim et al. 2001). The next major period of activity of
\sgr~came in 2001 (Guidorzi et al. 2001; Hurley et al.  2001a, b;
Feroci et al. 2001; Ricker et al. 2001a, b, c; Montanari et al. 2001,
Feroci et al. 2004).
During this episode, it became apparent that this source emits not
only the common, short SGR bursts, with durations of about 200 ms
and fluences  $\gtrsim 10^{-6} \rm erg \, cm^{-2}$, and
the much rarer giant flares, lasting for minutes and having fluences
$\gtrsim 10^{-3} \rm erg \, cm^{-2}$, but also, a class of
high-fluence \it intermediate \rm bursts, whose durations and fluences
fall somewhere in between, and may in fact form a continuum
(Kouveliotou et al. 2001; Woods et al. 2003; Feroci et
al. 2003).  Such events had been observed in the aftermath of the 1998
August 27 giant flare, but were thought to be related to it.  In
retrospect, it seems likely that such bursts were also emitted by
SGR0525-66, in the aftermath of the famous 1979 March 5 giant flare
(see Golenetskii et al. 1984). The properties of these intermediate
bursts are interesting for numerous reasons, not the least of which
are that some display X-ray afterglows, similar to the one observed
during the 1998 August 27 event (Feroci et al. 2001; Thompson and
Duncan et al. 2001), and that their longer duration permits 
studies of their spectral evolution.

Duncan and Thompson (1992), Paczy\'nski (1992), and Thompson and
Duncan (1995, 1996) have proposed that the soft gamma repeaters are
\it magnetars \rm, i.e., neutron stars with magnetic fields B $\rm
\approx 10^{15} G$.  In this model, magnetic dissipation causes the
neutron star crust to fracture, and Alfv\'en waves accelerate
electrons, resulting in short (200 ms) bursts of soft gamma-radiation.
Much more rarely, magnetic reconnection provides the energy for a
longer, extremely energetic giant flare involving the entire neutron
star magnetosphere.  The periodicity observed in both the quiescent
soft X-ray emission from \sgr~and in the giant flare, the high
spin-down rate, and the energetics of the giant flare, are all
consistent with the main features of the magnetar model.

In this paper, we analyze an intermediate burst from \sgr , which
occurred on 2001 July 2, emphasizing broad band 
and time-resolved X-ray spectral modeling
obtained using the data of the FREGATE (FREnch GAmma-ray TElescope) 
and WXM (Wide Field X-ray Monitor) experiments aboard
the HETE (High Energy Transient Explorer) spacecraft.

\section{Observations and Instrumentation}

\subsection{The FREGATE spectrometer} 
\subsubsection{Description}
\label{fregate}
The FREGATE gamma-ray burst (GRB) experiment has been described in detail (see Atteia et
al. 2003 for the details of the instrument and
data modes), so we will only summarize here its most relevant
characteristics.  It consists of four NaI(Tl) cleaved crystals, with
total on-axis area 160 cm$^2$, sensitive to photons in the energy
range 5-400 keV.  It has various modes for reading out data, including 
continuous 128
channel energy spectra with 5 second resolution, continuous 4
channel energy spectra with 0.16 s resolution,
and individually
time- and energy-tagged photons which are recorded when a burst trigger occurs.
This last data type, which allows detailed studies of the spectral
evolution of bright bursts, consists of 256000 photons (64000 per detector) in
256 energy channels spanning the range 5-400 keV whose times
are tagged to 6.4 $\mu$sec. In the present paper, we do not use data
below 7 keV because of instrumental noise.

The unique features of FREGATE as far as the present observations are
concerned are first, its wide energy range, from X-ray to gamma-ray
energies, and second, its good energy resolution, which ranges
from $\rm \sim 12 \% $ at 122 keV to $\rm \sim 42 \% $ at 6 keV.  The
wide energy range is particularly interesting for the observations
described here, because this is the first time that both the X-ray and
soft gamma-ray components of an intermediate SGR burst have been observed by a
single experiment.

\subsubsection{Calibration}
\label{fregcalib}
An account of the procedure used to construct the
response matrix is given in Olive et al. (2003).  The spectral
response of FREGATE at low energies, which is important for
the analysis presented here, has been tested in flight with the
Crab nebula and with the diffuse X-ray background (DXRB).  These
calibrations led to the discovery of an absorption which was larger
than expected below 15 keV (e.g. the absorption measured at 8 keV is 32\%
instead of the expected 20 \%).
This absorption has been attributed to a degradation of the
surface of the crystal, and corrected empirically in the response matrix.  
With this correction, FREGATE
measures the correct parameters for the spectra of the Crab 
(at angles ranging from 0 to 45 degrees) and for the DXRB.
In addition, the agreement between FREGATE and the WXM for bright GRBs
(for which we have enough statistics to do useful comparisons) is
better than 5\% in their common energy range, 7 to 20 keV.
We therefore believe that the FREGATE spectral
response is well known and reliable down to 7 keV at least.

The spectral analysis of very bright transients may also suffer
from dead time and pulse pile up effects in the detectors.
These effects were carefully measured on the ground before launch. 
The dead time in the burst buffers measured on the ground 
was found to be $ \sim 17 {\mathrm \mu s}$ (this is the {\it measured} dead time 
resulting from the combination of various electronic dead times).
For the brightest 25 ms of the burst the dead time was close to 20\% , so the true counting
rate at maximum could have reached 15000 cts/sec per detector above 5 keV.
During the decreasing part of the burst, the dead time was 
between 5\% and 10\%. 
The  FREGATE  electronics were specially designed to avoid spectral 
distorsion at high count rates. Even with significant dead time the spectra
are not distorted as long as pile up remains negligible. 
Pile up occurs when two photons arrive in a time window of 1.2 $\mu s$.
This means that pile up can affect at most 2\% of the counts at the
maximum of the peak, at most 1\% at the beginning of the interval following
the peak (interval I4, see below), and 
at most 0.5\% of the counts during most of the rest of burst.
We thus consider that the spectral analysis presented below is not
significantly affected by the burst high counting rate.

\subsection{The WXM spectro-imager} 
\subsubsection{Description}
\label{wxm}
The WXM is composed of two identical units each consisting of a 1 dimensional coded
mask and two 1 dimensional position-sensitive proportional counters (PSPCs) 
placed 187 mm below the mask.
The two units are placed in orthogonal directions.
The WXM covers the energy range 2-25 keV, with a maximum effective
area reaching 114 cm$^2$ at 8 keV, when the absorption of the mask
is taken into account.
The WXM has various data readout modes, and when a trigger occurs
individual photons are recorded, which are tagged in time, position 
and energy. These data were used in section \ref{spectral} to
construct the joint spectrum (WXM+FREGATE) of the burst
discussed here.
\subsubsection{Calibration}
\label{wxmcalib}
The energy response of the WXM has been constructed using an
analytical model of the response measured on the ground.
This model takes into account the mask pattern, the absorption
coefficients for the gas in the PSPCs, the energy-channel conversion 
and the energy resolution measured within the volume of the detector.
The response of the WXM has been tested in flight with the Crab nebula,
measured at angles varying from 0$^\circ$ to 30$^\circ$ off-axis.
The spectral slope is correctly measured, as well as n$_H$, the column
density of the gas, except at large incident angles ($\sim 30^\circ$)
where n$_H$ appears to be underestimated. This last feature may be 
relevant for the burst studied here, which arrived at an angle of 
32.7$^\circ$.
For bursts with count rates above $\sim$ 4000 c/s, the WXM experiences 
significant dead time.
In the present case, The dead time in the WXM varies significantly 
during the burst, resulting in a WXM light curve which does not 
closely follow the light curve recorded by FREGATE at low energies 
(see Fig. 2 of Torii et al. 2003).
In addition the housekeeping data needed to measure the dead time 
of the WXM were only available on a 5~s timescale for this burst. 
These two factors have the consequence that
we can use the WXM data only for the study of the time averaged spectrum
of the July 2 burst (Section \ref{spectral}). 
The time-resolved analysis (Section \ref{timeresolved}) must be
based on FREGATE observations only.

\subsection{Context: the 2001 activation} 

The 2001 activation of SGR1900+14 began on April 18 and ended around
July 8.  During that period it emitted over 100 bursts.  The exact
number is uncertain for two reasons.  First, many events were detected
by a single instrument in the IPN and could not be localized.  Second,
the source is known to emit weak and/or soft spectrum events (Woods et
al. 2001, 2003), and these would have been undetectable to the
experiments in the IPN.  However, among the events whose origin is
definitely SGR1900+14, two are noteworthy for their long durations and
high fluences: that of April 18 (Guidorzi et al. 2001, Hurley et al.,
2001a, Feroci et al. 2003), and the present 2001 July 2  event.

This burst (HETE trigger 1576) was detected by the FREGATE experiment at 12846.529 s UT
and also by the Wide Field X-Ray Monitor aboard HETE, and both
automated and ground-processed localizations were sent out. The WXM
error circle, which has a radius of 12 \arcmin, includes the position
of \sgr~(Ricker et al. 2001c).  In addition, this event was observed
by a number of spacecraft in the IPN: \it Ulysses, Mars Odyssey \rm
(HEND experiment) \it BeppoSAX \rm (GRBM experiment), and \it Wind \rm
(Konus experiment). The triangulated position is also in agreement
with that of \sgr (Hurley et al. 2001a). 
At the time of the burst, the HETE detector axis
was pointed at $\rm \alpha(2000)=280.4^o, \, \delta(2000)=-22.8^o $,
32.7 degrees from SGR1900+14.  Thus the effective geometrical area 
exposed to SGR 1900+14 was 135 cm$^2$ for the detectors of FREGATE 
and 14.6 cm$^2$ for the WXM detectors.

\section{Temporal analysis}
\label{tempo}

\begin{figure}
\plotone{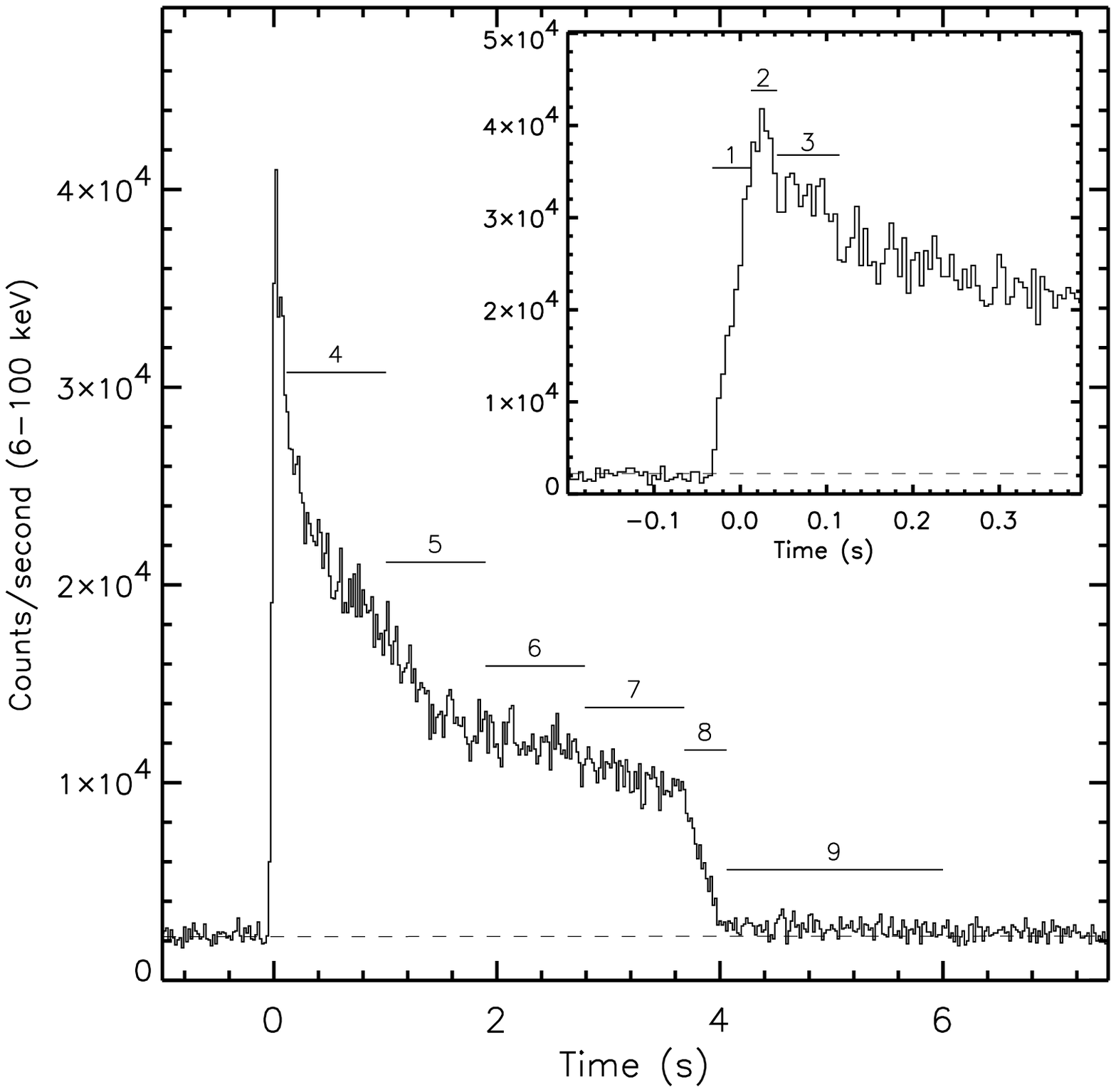}
\caption{The 7-100 keV time history of the 2001 July 2 burst from SGR1900+14 
as observed by FREGATE
(time resolution 20 ms). Inset: initial part of the burst
(time resolution 5 ms).  The interpolated background is indicated by
a dashed line.  The start time corresponds to 12846.529 UTC (HETE
trigger time). Horizontal lines have been drawn and labeled to
indicate the intervals corresponding to the 9 spectra of figures
\ref{f3} and \ref{f3b} (I1 to I9). For the detailed time-resolved spectral analysis,
intervals I4 to I7 have been further subdivided into 19 sub-intervals (not shown).
\label{f1}}
\end{figure}

The 7-100 keV time history of the 2001 July 2 burst from the
time-tagged FREGATE data is shown in figure
\ref{f1} where the reference time is the HETE trigger time ($T_{tr}$).
FREGATE recorded a total of $1.07 \times 10^5$ counts from the burst.
At maximum the total count rate on a 25 ms timescale reached 43000 cts/sec.


Using the continuous 0.16 s, 4 channel FREGATE data and the
WXM data,
we have searched for evidence of SGR activity before
(i.e. precursors) and after (i.e. extended tail) the
main peak which triggered FREGATE. With the possible exception of a
very weak tail, discussed briefly in section 4, no such
activity was found. 
The sensitivity of FREGATE for a 1 second long signal with an SGR 
spectrum is of the order of $4 \times 10^{-8}$ erg cm$^{-2}$ s$^{-1}$. 
This sensitivity guarantees the detection of a precursor 1.5 times fainter
than the one observed by Ibrahim et al. (2001) before the burst of August 29th.
It is however, much too low to permit the detection of a tail
which would be comparable to the tail of the August 29th burst.
The instrumental background, plotted as a dotted line in
figure \ref{f1}, was interpolated using the count rates
prior to $T_{tr}-2$ s and after $T_{tr}+6$ s. The total duration
of this intermediate burst is $\rm T_{90}= 3470 \, ms $; this
may be compared to a typical SGR burst
duration of $\sim$ 200 ms.

Motivated by the light curve morphology, we divided the burst into
9 consecutive and contiguous intervals (hereafter designated I1 to I9).  The burst
begins with a sudden rise in the count rate (I1, lasting 45 ms, with an
e-folding rise time of 32 ms), followed by an
initial spike (I2, with a duration of 30 ms) and a plateau (I3, with a duration of 70 ms). This
is followed by a smooth decay with an e-folding time of 3.9 s, lasting $\sim$ 3400 ms, that we
have divided into 4 equally long intervals (I4 to I7), and a final, faster decay lasting
300 ms (I8, duration 380 ms)\footnote{Fig. 2 of Olive et al. (2003) was erroneous.
It showed a spike just before the decay which was an artefact due to 
an uncorrected timing error in the time-tagged photon data.}. 
The event ends with a faint $\sim$ 2s long decreasing tail (I9).




\section{The 2-150 keV time-integrated spectrum}
\label{spectral}
\subsection{Spectral modeling}
\label{modeling}

We have used XSPEC (Arnaud 1996) to deconvolve the WXM and FREGATE count
spectra. In all our spectral analyses we have taken into account a Galactic absorption 
n$_H$ = $2.5 \times 10^{22}$ cm$^{-2}$, determined from the ASCA measurements
of the quiescent X-ray counterpart (Hurley et al. 1999b).
Since there are no source counts in the spectrum above 150 keV, we
first attempted to fit the 2-150 keV time-integrated count
spectrum (taken over I1 to I8) using several single component
models. We tried a single temperature optically thin thermal
bremsstrahlung spectrum (OTTB), a simple or broken power law (PL, BKPL)
and a single blackbody (BB).  None of these models reproduces
the shape of the measured spectrum from
$\rm$ 2 to $\rm$ 150 keV, and all of them may be rejected on statistical grounds
($\chi^{2}$/dof= 2146/95, 19135/95, 890/93, 1965/95 respectively). 
A similar conclusion applies to the
time resolved spectra of intervals I1 to I9 taken separately.
The OTTB function has been successfully used to fit the spectra of many 
of the short SGR bursts above 15 keV.
In the present case it fails to reproduce the rollover of the spectrum 
observed by FREGATE below 15 keV (an illustration of this failure
can be found in Fig. 3 of Olive et al. 2003).
We note that a similar rollover in the spectrum of bright 
SGR bursts, was previously reported by Fenimore et al. (1994), 
based on observations of SGR 1806-20 with the {\it International Cometary Explorer}.
More recently, Feroci et al. (2004) also reported a deficit of 
photons below 15 keV in 10 short bursts from SGR 1900+14 observed
with the \it BeppoSAX \rm narrow field instruments.

As an alternative, we have tried the magnetar spectrum of Perna et
al. (2001). In this model, the spectrum originates as thermal emission
from the neutron star surface, but includes the effects of anisotropic
heat flow through a magnetized neutron star envelope, reprocessing in
a light-element atmosphere, and relativistic corrections.  R. Perna
has kindly extended her results to the higher energy range of FREGATE
to make this possible (private communication, 2002). However, we do not
obtain an improved fit using it.

\begin{figure}
\epsscale{0.7}
\plotone{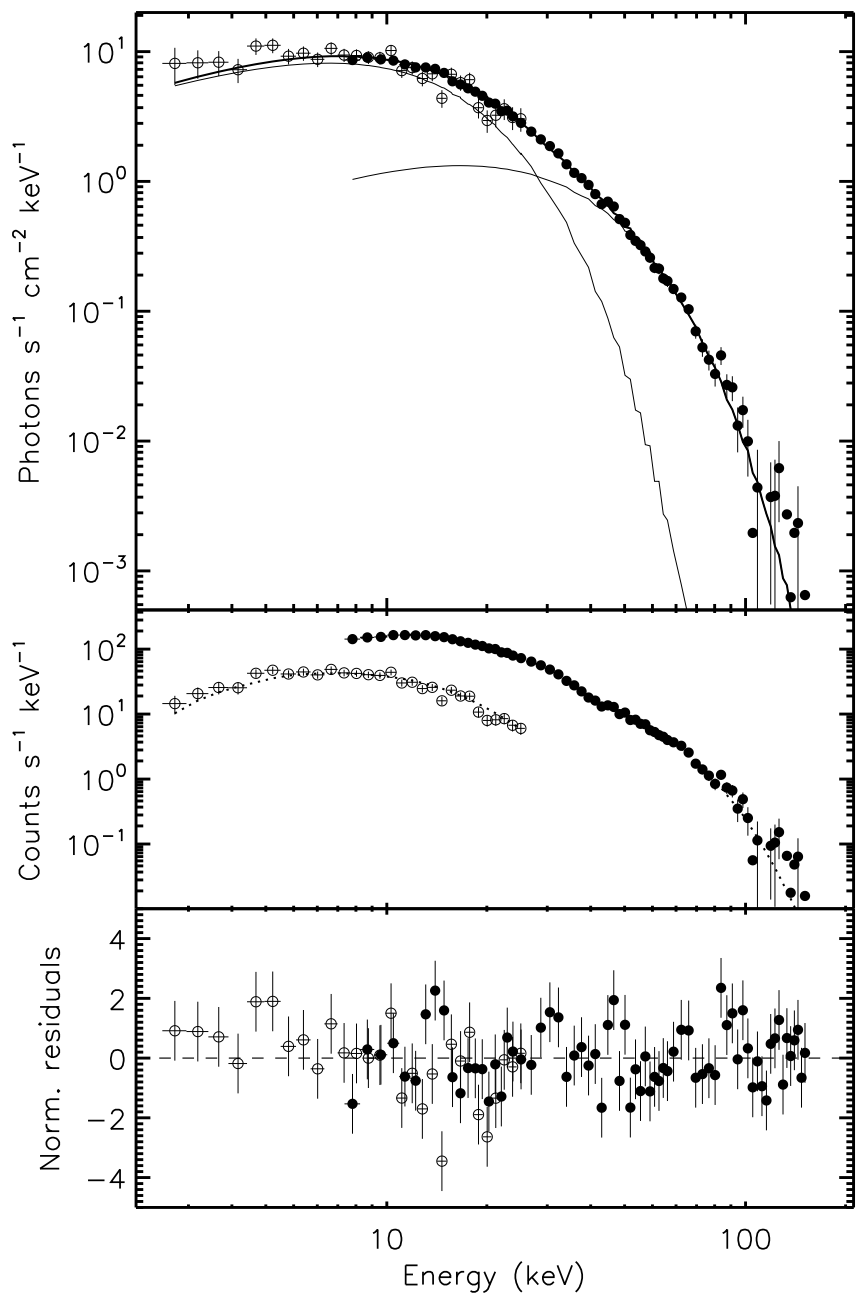}
\caption{
Top: Deconvolved (photon) spectrum of the time-integrated
burst (I1 to I8, see figure \ref{f1}) using a two blackbody model. 
The two blackbodies and their their sum are plotted.
The filled circles correspond to the FREGATE 7-150 keV
spectrum and the open circles to the WXM 2-25 keV spectrum.  
Middle: Observed (count) spectrum of the time-integrated burst and 
the best fit model (dotted line).
Bottom: Residuals between the data and the model normalized to the
standard deviation. 
\label{f2}}
\end{figure}

Next we tried several two-component models, which included combinations
of PL,
BKPL, and OTTB. They also failed to reproduce the shape of the spectrum: the
best fit was given by a BKPL+BKPL model, which is however
clearly rejected ($\chi^{2}$/dof= 358/89).  We eventually found that the only
two-component model that can be fit to the experimental spectrum is
composed of the sum of two blackbodies (BB) with different temperatures. 
Hereafter we will refer to these two thermal components
as the Low and the High Temperature
Blackbodies (LTBB and HTBB respectively). The experimental spectrum
and its residuals are plotted in figure \ref{f2}. The
temperatures for this time-integrated spectrum are
kT$_{\mathrm LTBB} = 4.25 \pm 0.1$ keV and kT$_{\mathrm HTBB} = 9.8 \pm 0.3$ keV.

The reduced $\chi^{2}$ value that we derive for this fit
($\chi^{2}$/dof= 137/93, corresponding to a null hypothesis
probability of 0.002) is still not completely satisfactory.
However, the $\chi^{2}$ value can be reduced to $\chi^{2}$/dof= 104/93 
(null hypothesis probability = 0.2)
by adding a 3\% systematic error in quadrature to the
statistical errors to account for calibration uncertainties
affecting this high signal spectrum.
Thus we conclude that the HTBB+LTBB model provides an acceptable fit 
to the measured WXM+FREGATE spectrum over the 2 to 150 keV energy range.

\subsection{Two vs. three blackbodies}
\label{2ou+} 
Since the temperatures of the two blackbodies in our fit differ by not much more than 
a factor of two, the two blackbodies are not detected individually, and this
raises the possibility that
the data could favor a model 
based on a sum of blackbodies with a narrow distribution of temperatures.   
To answer this question we tried to fit the time-integrated spectrum 
with a three-blackbody model (the three BB model intending to 
simulate a narrow distribution of temperatures).


The three-blackbody model gives a good fit with temperatures 
equal to 1.4, 4.4, and 10.0 keV, and corresponding radii of 55, 23, and 4.0 km
(at 10 kpc).
The fit is fully acceptable ($\chi^{2} = 81$ for 91 dof), and provides
some improvement over the two-blackbody model ($\chi^2 = 104$ for 93 dof).
A close examination of the three BB fit shows that the additional blackbody component
(with kT=1.4 keV) is used to fit the few points below 5 keV 
which are above the two BB fit (see figure \ref{f2}).
Since the spectral response of the WXM at large incident angles
is subject to larger uncertainties (section \ref{wxmcalib}) and since
the two BB fit is statistically acceptable (section \ref{spectral}),
we conclude that, to within the limits imposed by the data, 
a third blackbody component is not strongly required, 
and we
restrict the following discussion to the two-blackbody model.


\subsection{Burst energetics}
\label{energetics}
The spectral analysis was used to derive the burst energetics. A
2-150 keV fluence 
of $\rm 19 \times 10^{-6}$ erg cm$^{-2}$ and a
$\sim$ 3.5 s duration place this event firmly in
the ``intermediate'' class\footnote{The values given here are the ones
{\it measured} with FREGATE. Note that, considering the amount of dead time 
(see section \ref{fregcalib}), the fluence given here should be increased by 10\% to 
obtain the true fluence.}. Specifically, this event looks similar
to other intermediate bursts from SGR 1900+14 described in the literature, 
such as 981028b in Aptekar et al. (2001) (duration = 4 s, 
fluence= $\rm 48 \times 10^{-6}$ erg cm$^{-2}$), and 
the  ``unusual burst'' extensively discussed by Ibrahim et al. (2001) 
which had a duration of 3.5 seconds and a fluence above 25 keV 
of $\rm 19 \times 10^{-6}$ erg cm$^{-2}$
(vs. $\rm 7.8 \times 10^{-6}$ erg cm$^{-2}$ for the event discussed here).


Assuming for the sake of concreteness a distance of 10
kpc, the average luminosity in the 2-150 keV energy range 
is $ \sim \rm 6.0 \times
10^{40}$ erg s$^{-1}$ and the energy release is $ \rm 2.1 \times
10^{41}$ erg.



\section{Comparison with previous work}

\subsection{Spectral analyses of intermediate bursts}

Detailed spectral analyses are available for two other intermediate bursts
from \sgr : the 1998 August 29 event observed with BATSE and RXTE, and the 
2001 April 18 event observed with the BeppoSAX Gamma-Ray Burst Monitor (GRBM). 
It should be noted that FREGATE covers energies below 
25 keV, which were not covered by BATSE or the GRBM, but that the FREGATE
data do not constrain
the burst spectrum above 300 keV, due to decreasing sensitivity at these
energies. 

Ibrahim et al. (2001) found that the spectrum of the 2001 August 29 burst 
was correctly fitted with an OTTB (kT=20.6 keV) {\it above 25 keV}.
We find that an OTTB is also a reasonable fit to the 2001 July 2 burst spectrum
above 25 keV (\chisq /dof = 57.2/47) with kT = 23.0$^{+0.67}_{-0.57}$ keV. 
Thus the results reported in this paper are consistent with those
of Ibrahim et al. (2001); indeed, it is possible that, had $<$25 keV data
for that event been available, a two blackbody model might have been
preferred for it, too.

Overall, the burst of 2001 April 18 is longer ($\sim 40$ s) and 
more energetic (with a 40-300 keV fluence $\sim \rm 120 \times 10^{-6}$ erg cm$^{-2}$) than
the event discussed in this paper.
Guidorzi et al. find that the spectrum of their bright burst
is well fitted by a combination of three spectral components (see
their Table 2): a 14 keV blackbody, a broken power law
with a break at 73 keV, and a single power law with an index close to $-0.5$.
Replacing the blackbody by a 32 keV OTTB
gives an equally good fit.
The single power law dominates the spectrum at energies above 300 keV, but
we will not discuss it here because it is outside the FREGATE energy range.
We tried to fit the spectrum of our event with a model which included
a blackbody and a broken power law. The best fit is obtained with
a blackbody of temperature 5.1 $\pm 0.2$ keV plus a broken power law
with a break at 53.5 $\pm 5$ keV.  However, the reduced \chisq\ is 
1.75 (\chisq = 112 for 64 dof), indicating that this fit is not acceptable.
While the FREGATE burst cannot be fit with the sum of a blackbody and a broken power law,
Guidorzi et al. mention that a fit with the sum of two blackbodies
cannot be ruled out for the burst of 2001 April 18.
C. Guidorzi kindly provided us 
the best fit parameters of their two blackbody fit
in the energy range 40-300 keV (private communication, 2003). Their temperatures are
13.6 (-0.2;+0.1) keV and 35 (-3;+2) keV, with \chisq = 72.6 for 66 dof.
While their low temperature blackbody could be compared with our high
temperature blackbody, we found no evidence in the FREGATE data for a 35 keV blackbody. 
We tried to fit our 40-300 keV spectrum (corresponding to the energy range
of the Guidorzi et al. analysis) with two BBs with temperatures 
of 13 keV and 35 keV, but found that convergence was 
attained for two blackbodies
having temperatures of 4.1 keV and 10.6 keV (\chisq = 53.0 for 66 dof).
If we freeze the temperatures of the BBs to the values obtained by 
Guidorzi et al. we obtain a \chisq\ of 568 (for 68 dof), and the
35 keV BB has a normalization of 10$^{-7}$, indicating 
that it is rejected by the data.
We therefore conclude that 
the intermediate bursts of 2001 July 2  
and 2001 April 18 have very different energy spectra.

\subsection{Spectral analyses of short bursts}
In parallel with the analysis of the intermediate burst reported here, 
we have performed the spectral analysis of several classical short
bursts. We find that the two BB model adequately fits the spectrum
of short SGR bursts with parameters similar to those given in Table \ref{tab1}. 
A more detailed discussion of the spectral modeling of short
SGR bursts is in progress and will be reported in a 
future paper (Atteia et al. in preparation).
Recently Feroci et al. (2004) presented broad band spectra of ten short bursts
from SGR 1900+14 observed with the narrow field instruments of BeppoSAX. 
They find that the OTTB model, which successfully fits the spectrum of
short SGR bursts above 15 keV, is not acceptable below 15 keV.
They also find that a two BB or a cutoff powerlaw model provide good 
spectral fits to the short SGR bursts in the energy range 1.5 to 100 keV. 
With the two BB model, the ten bursts are jointly fit
with a model having temperatures at 3.3 keV and 9.5 keV, similar to the
temperatures reported in our Table \ref{tab1}.

Following Feroci et al. (2004), we tried to fit the WXM+FREGATE spectrum with a cutoff powerlaw 
model and find it to be excluded by the data ($\chi^2 = 170$ for 94 dof,
with 3\% systematic error, vs. 104 for 93 dof for the two BB model). 
The corresponding null hypothesis probability is $2 \times 10^{-6}$ vs. 0.2
for the two BB model.

\section{Time-resolved analysis}
\label{timeresolved}


In the first part of this section we display the spectra of intervals I1 to I9,
illustrating the validity of the two BB model on short timescales
(section \ref{shortspectra}).
In a second part (section \ref{evolpara}) we study the evolution 
of the model parameters in 24 sub-intervals, chosen to provide
the finest temporal resolution while keeping enough counts
to perform a statistically significant analysis.

\subsection{Time-resolved spectra}
\label{shortspectra}

\begin{figure}
\epsscale{1.1}
\plotone{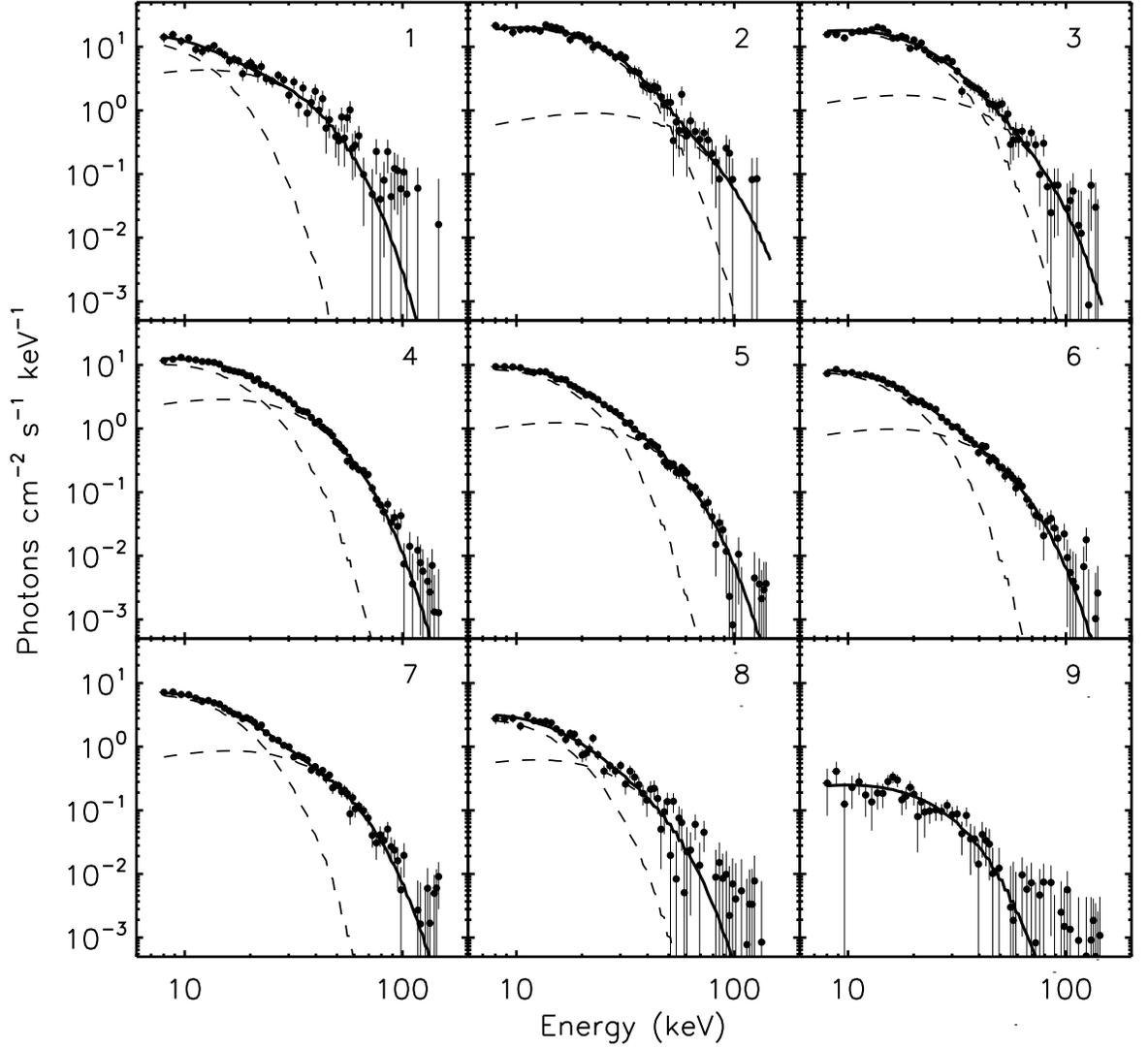}
\caption{FREGATE deconvolved (photon) spectra and the best fit model 
(the sum of 2 blackbodies, shown as dashed lines, and their sum, shown
as a solid line) for the 9 intervals of the
burst of 2001 July 2 indicated in
Figure \ref{f1} (I1 to I9). Interval I9 can be fit with a single
blackbody. 
The spectral parameters for these 9 spectra are given in Table \ref{tab1}.
\label{f3}}
\end{figure}

\begin{figure}
\epsscale{1.1}
\plotone{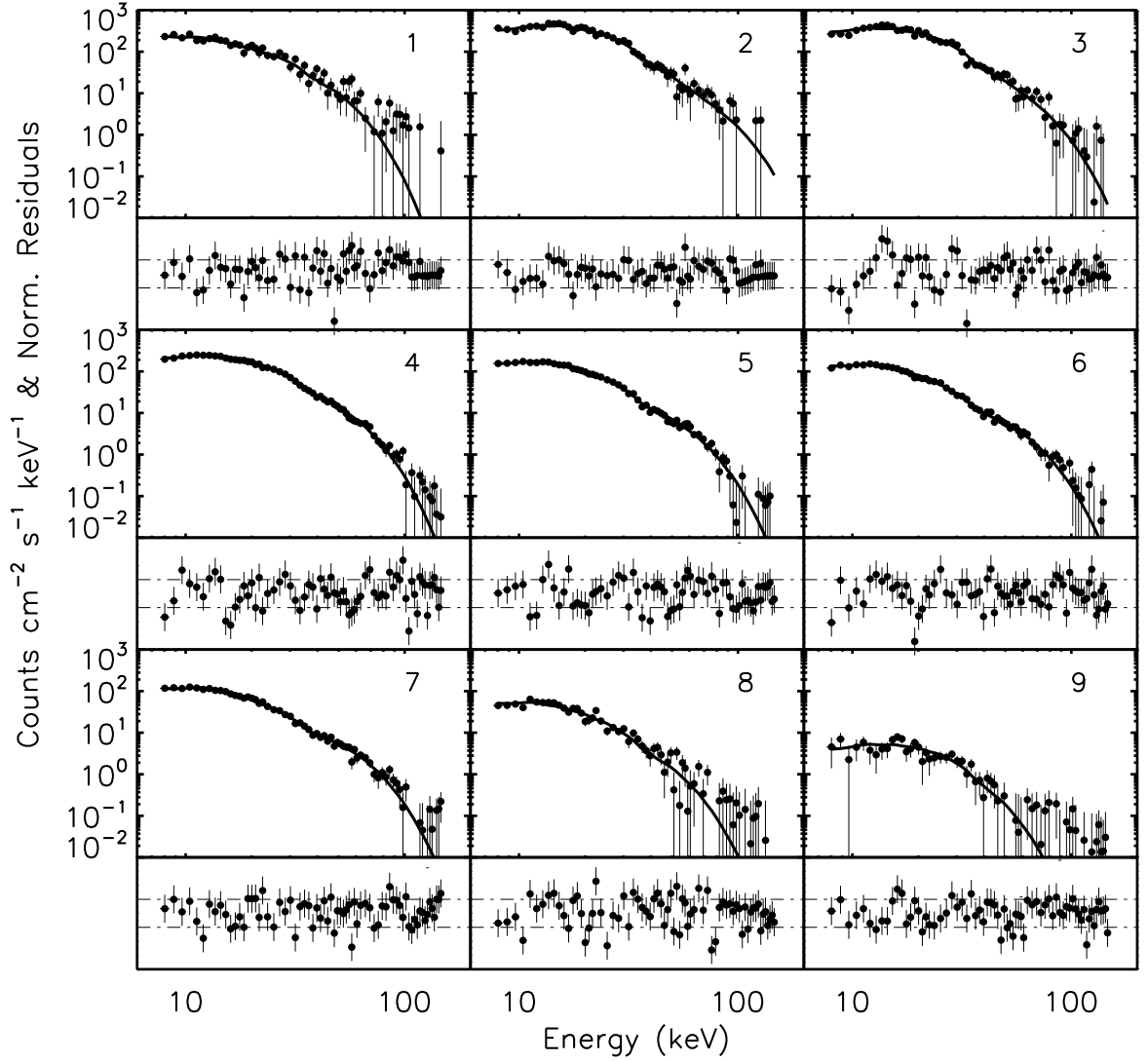}
\caption{FREGATE count spectra and residuals for the two blackbody model 
for the 9 intervals of the burst of 2001 July 2 indicated in figure \ref{f1}.
The two horizontal dashed lines indicate $\pm$ 1 sigma confidence limits on the residuals. 
The last interval can be fit with a single
blackbody. The spectral parameters are given in Table \ref{tab1}.
\label{f3b}}
\end{figure}

To approximate the instantaneous spectral shape,
we have fit the spectra of intervals I1 to I9 (figure \ref{f1}) with a two BB model.
As explained in section \ref{wxmcalib}, this study can only
be done with FREGATE, and therefore it is restricted to the energy range 7-150 keV.
We successfully fit the spectra of I1 to I8 with the two BB model,
while I9 could be fit with a single BB component.
Table \ref{tab1} summarizes the results.  
Here kT$_{LTBB}$ and kT$_{HTBB}$
are the temperatures of the two blackbodies, R$_{LTBB}$ and R$_{HTBB}$
are the radii of the emitting regions, and L$_{LTBB}$ and L$_{HTBB}$ are bolometric
luminosities in units of 10$^{40}$ erg s$^{-1}$ 
for an assumed distance of 10 kpc.  
In figure \ref{f3}, we have plotted the 9 unfolded spectra and compared them to the
model (the spectrum of I9 was fit with a single blackbody model). 
For all spectra, the fit is fully acceptable, the worst case
being for I3 ($\chi^{2}$/dof = 81.5/66, corresponding to a null
hypothesis probability of 0.1).  
In figure \ref{f3b}, we have plotted the 9 count spectra and the residuals
of the fit.
Table \ref{tab1} and figure \ref{f3} show that 
i) the two blackbody fit works on short timescales, and 
ii) the temperatures of the two blackbodies and the overall 
spectral shape evolve significantly during the burst.

\subsection{Temporal evolution of the spectral parameters}
\label{evolpara}

In order to study accurately the variations of the spectral parameters 
along the burst, we have sub-divided this event into 24 intervals, 
noted i1 to i24 and corresponding to the former I1, I2, I3, I4 to I7 
divided into 19 sub-intervals, I8, and I9.  These intervals are the
shortest ones that can be selected with good statistics.

\begin{figure}
\epsscale{1.1}
\plotone{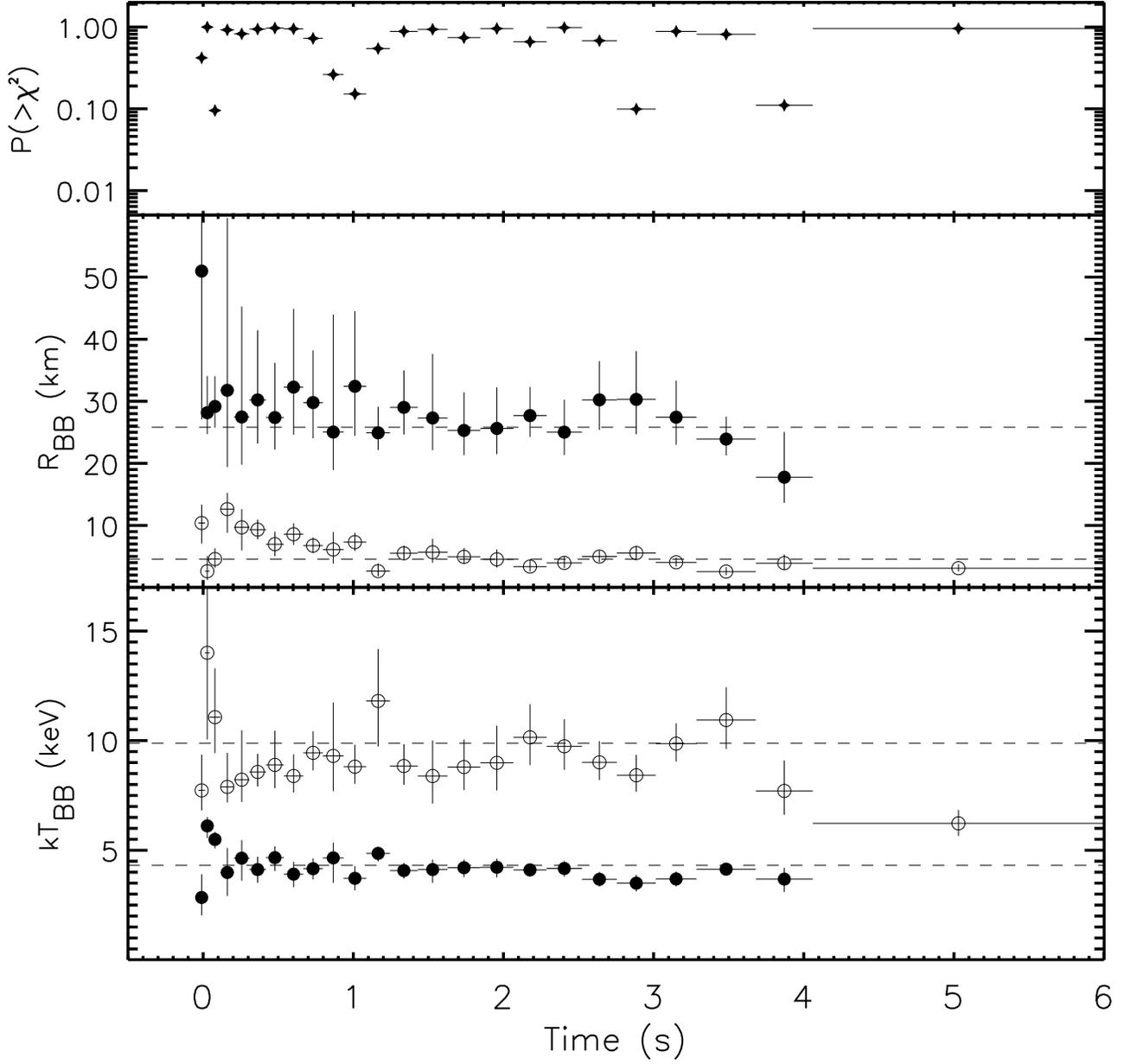}
\vspace{0.5in}
\caption{Spectral parameters of the two BB best fit as a function of time
for 24 time intervals i1 to i24 (see text).  The bottom panel displays
the temperatures of the LTBB (filled circles) and HTBB (open circles).
The middle panel displays the radii of the LTBB (filled circles) and HTBB (open circles),
for a source distance of 10 kpc. Note the decreasing trend of the
radius of the HTBB  The error bars are for 90\% confidence level
(defined as the parameter interval for which the $\chi^{2}$ is less than $\chi^{2}_{min}+2.7$
for variation of a single parameter, re-fitting the spectrum at each step). 
The upper panel shows the null hypothesis probability of the fit, 
and indicates that all the fits are acceptable. 
\label{f4}}
\end{figure}

\begin{figure}
\epsscale{0.6}
\plotone{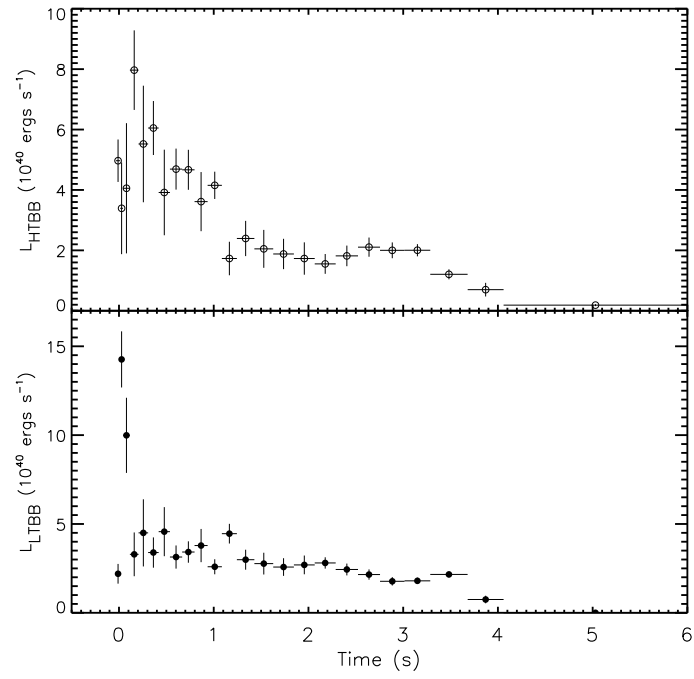}
\vspace{0.5in}
\caption{Bolometric luminosities of the LTBB and HTBB as a function
of time, for the 24 fits (i1 to i24, see Table \ref{tab2})
\label{f5}}
\end{figure}

We have applied the fitting procedure with the LTBB+HTBB model 
to the resulting 24 spectra. 
The derived parameters (blackbody radius and temperature) 
are plotted as a function of time in figure
\ref{f4} and reported in Table \ref{tab2}. 
In figure \ref{f4}, the dashed horizontal lines indicate the value of the 
parameters measured for the time-integrated spectrum. 
The luminosities of the two components as a function of time
are plotted in figure \ref{f5}. 
A crude analysis of the variability of the four spectral parameters
shows that the two temperatures, as well as the radius of the low temperature 
blackbody, display comparable fluctuations : $\sigma(log(T_{\rm LTBB})) = 0.068$, 
$\sigma(log(T_{\rm HTBB})) = 0.061$, and $\sigma(log(R_{\rm LTBB})) = 0.068$.
There is no statistically significant evidence that either temperature or
R$_{\rm LTBB}$ evolves with time. 
The radius of the high temperature blackbody, on the other hand, shows significantly 
larger fluctuations, with $\sigma(log(R_{\rm HTBB})) = 0.197$, and it tends to
decrease with time.

\begin{figure}
\epsscale{1.1}
\plotone{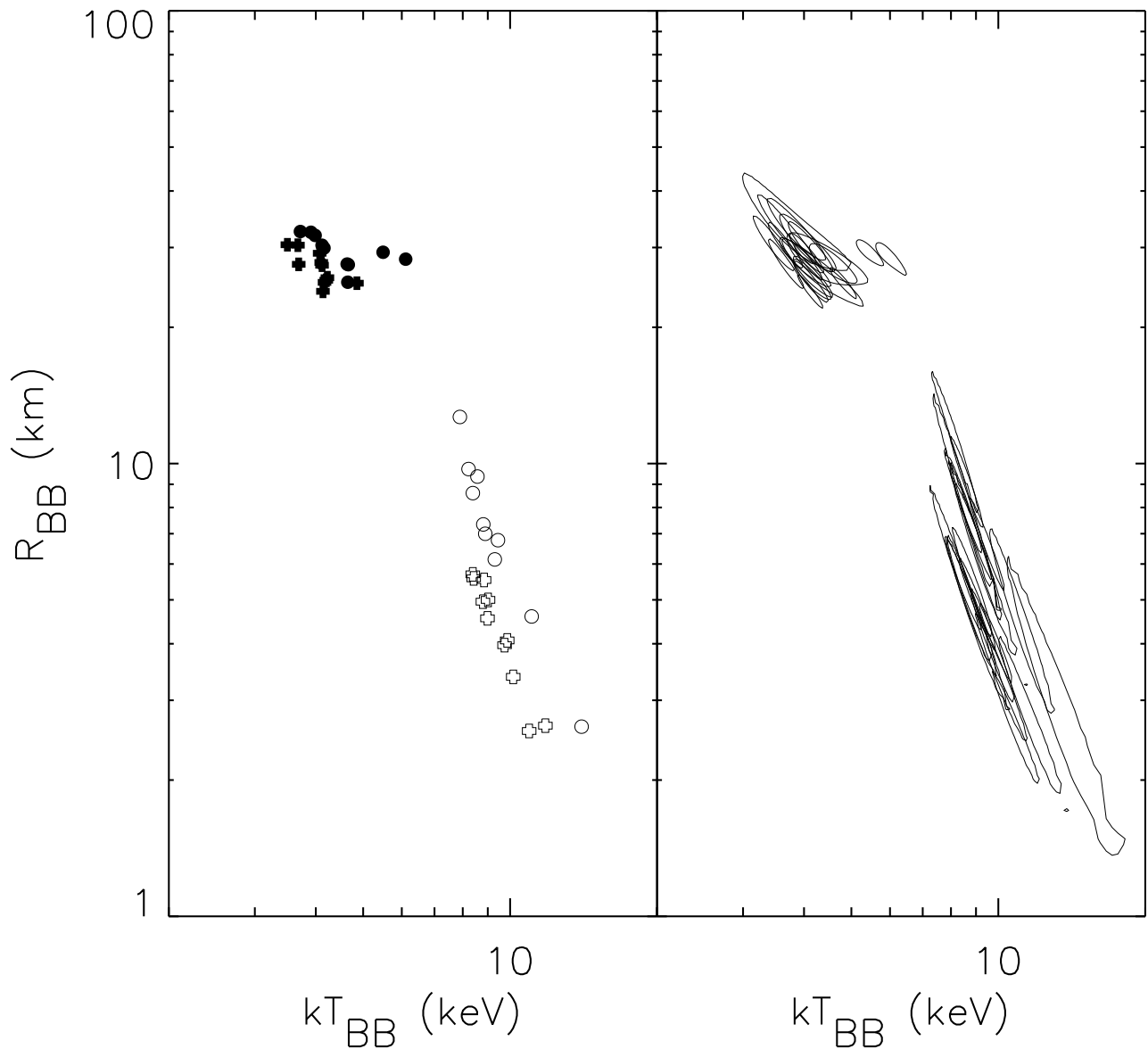}
\caption{Left panel: The radii of the two blackbodies as a function of their temperatures,
for intervals i2 to i22 (see Table \ref{tab2}). 
Filled symbols are used for the LTBB and open symbols for the HTBB.
The circles show the parameters for intervals i2 to i11, and the crosses for 
intervals i12 to i22. 
Right panel: Contour plot of the 90 \%  confidence region ellipses for each data point.  
Note that the radius of the LTBB (filled symbols) does not vary with the temperature, 
while the radius of the HTBB (open symbols) is anti-correlated with the temperature. 
\label{f6}}
\end{figure}

In order to test for possible correlations between the fit parameters
we have plotted in figure \ref{f6} the 
temperature as a function of the radius for the two blackbodies, 
for the 24 intervals except intervals i1, i23, and i24 whose
parameters are not well constrained. 
The two blackbodies exhibit different emission patterns: 
while the radius of the low temperature
blackbody is nearly constant and independent of the temperature, 
the high temperature blackbody shows a clear anti-correlation between
the radius and the temperature. A correlation analysis using the Pearson
test shows that this anticorrelation is significant at the 4.4 $\sigma$ level. 
The shapes of the ellipse errors, however, indicate that
the observed correlation is at least partly due to the fitting procedure.
Indeed, the parameter relationship in the blackbody model is strongly non-linear.

\section{Discussion}
\label{discussion}

\subsection{Temperatures, radii, and luminosities}
\label{paraBB}

A first look at the parameters given in Table \ref{tab1} 
calls for the following remarks.
The temperature of the HTBB is very close to the typical 
temperature of a fireball trapped at the surface of a magnetar 
($\sim$11 keV) {\it predicted} by Thompson \& Duncan (2001).
The radii of the HTBB and LTBB are consistent with those 
of neutron stars and/or their magnetospheres. 
For distances less than or greater than the assumed 10 kpc, the radii would
scale linearly with the distance.
Finally, the time-integrated luminosities L$_{LTBB}$ and L$_{HTBB}$ 
emitted by the  
low temperature and high temperature components differ by less than 
20\%.\footnote{While the integrated luminosities are comparable, figure 
7 and table 1 show that the quasi-equality of the luminosities of the two 
blackbodies does not hold on short timescales, particularly at early times.}

\subsection{Comments on the spectral modeling} 
\label{interpretation}

We have shown that a two BB model fits the data. This model 
can be interpreted either as a physical or a phenomenological
description of the data.
In the first case, the model, taken literally, implies that
we are observing two distinct radiating volumes,
and the measured parameters give us insight into the physics 
of these regions.
In the second case, the fit is only a convenient 
representation of the emission and little or no physical meaning can
be attributed to the measured parameters.
At this stage, we believe that the study of a single event
is not sufficient to reach a firm conclusion about the physical meaning of 
the 2 BB description.
We expect that the study of a large number of SGR bursts
will eventually resolve this uncertainty, and 
in the meantime we discuss below the implications 
of these two possible interpretations. 

\subsection{The two blackbody interpretation}

\subsubsection{The HTBB: A trapped fireball ?}

Two observations lead us to explore here the hypothesis
that the HTBB originates from a fireball trapped on the surface
of the neutron star, as described in Thompson \& Duncan (2001).
The first is the remarkable agreement of the temperature of the HTBB 
inferred from the FREGATE data with the temperature predicted theoretically for 
a trapped fireball (9.8 vs. 11 keV). The second is the contracting radius
of the HTBB blackbody which is invoked in the trapped fireball model.

\begin{figure}
\epsscale{1.1}
\plotone{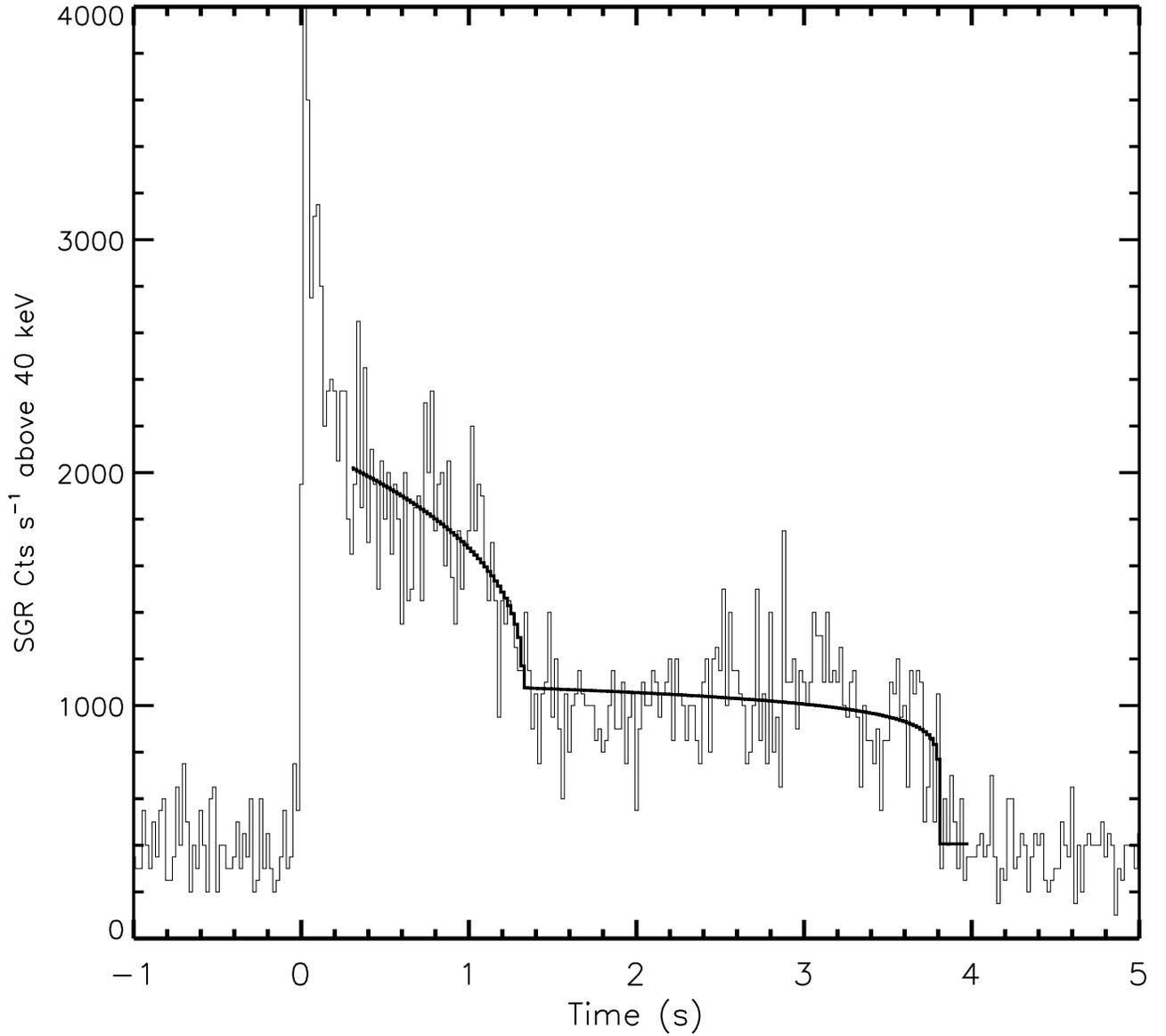}
\caption{Fit of the light curve above 40 keV with the sum 
of two trapped fireballs having evaporating times of 1.3 and 3.8 seconds 
and dimensions $\chi=0.4$ and $\chi=0.1$.  The time resolution is 20 ms.
The first 0.25 seconds of the burst (the first spike) are not 
included in the fit.
\label{f7}}
\end{figure}

Can a trapped fireball explain the temporal evolution of the HTBB emission?
The abrupt disappearance of the emission in about 0.3 s (figure
\ref{f1}, interval I8) is too rapid to be due to the rotation of the neutron star,
but it is consistent with the behavior of a trapped fireball, and we believe
that it reflects the intrinsic evolution of the emission.
To check this hypothesis we tried to fit the light curve above 40 keV, 
where the HTBB is dominant, with a trapped fireball model following
the temporal evolution described in Thompson \& Duncan (2001).
The light curve and the fit are displayed in figure \ref{f7}.
The light curve exhibits two episodes of fast decline which 
cannot be fit with a single trapped fireball. 
It is nicely fit, however, with the sum of two trapped fireballs
having evaporation times of 1.3 sec and 3.8 seconds, and dimensions
$\chi = 0.4$ and $\chi = 0.1$ (see Thompson \& Duncan 2001).
For comparison, $\chi = 2$ for a spherical fireball, and $\chi \sim 4$
for the 1998 August 27 event (Feroci et al. 2001).

We also note that we do not observe any modulation
of the emission with the 5.16 s period of SGR1900+14, even though the
burst is significantly longer than half the rotation period.  
The fact that the 5.16 sec  modulation has been observed many times
(in the quiescent soft X-ray emission and during the giant flare), 
indicates that the rotation axis of the magnetar is not directed towards the Earth. 
It is still possible, however, to have it not too far from the 
line of sight, allowing large 
regions to remain permanently visible and unmodulated.
If this burst took place in the polar region continuously
visible from Earth, this would explain the lack of rotational modulation.  

\subsubsection{The LTBB: thermal emission from the heated neutron star ?}

While the energy radiated by the two blackbodies is roughly equal,
their radii and temperatures are clearly different.
The size of the region emitting the LTBB is about 
30 times larger than the average size of the ``fireball''.
One possible explanation is that part of the 
fireball energy is reprocessed in a larger region and emitted at a lower temperature.
This would also explain the quick drop in luminosity
of the LTBB, a characteristic which it shares with the HTBB.
Under this hypothesis, the stability of the radius 
of the LTBB blackbody shows that the ``reprocessing region'' 
is not strongly affected by the amount of energy coming from the fireball.

Could the LTBB be emitted by the surface of the neutron star?
We check below whether the size of the region emitting the LTBB 
is compatible with the typical surface area of a neutron star.
Under this hypothesis, the LTBB spectral parameters (equivalent radius and temperature) 
are affected
by the gravitational redshift of the neutron star. 
For a given source distance, one can derive 
a relation between the {\it true} radius and the gravitational mass of the star.
These relations are plotted in figure \ref{f9}.
For example, at a distance of 10 kpc, a neutron star with a mass of 1.4 M$_\odot$ 
has a radius of 23 km.  
Distances of  4, 6, 8, 12, and 14 kpc
give radii of 5, 12.5, 18, 29, and 34 km respectively.
Given the uncertainty in the distance of SGR 1900+14 (see Section \ref{intro}),
our observations do not exclude the possibility that the LTBB is emitted
by the surface of the neutron star, but
the numbers above clearly show that the whole surface must radiate,
a fact which could help explain the absence of rotational modulation.

\subsection{The 'magnetar spectrum' interpretation}

Radiation transfer in the strong magnetic fields of magnetars has
been studied in some detail in recent years.
Here we discuss the interpretation of our data in terms of those theoretical
models of SGR bursts  which take into account the detailed radiation transfer
in superstrong magnetic fields.
This is mainly based on the work of Ulmer (1994)
and Lyubarsky (2002) which describes the expected X-ray spectrum of short SGR bursts. 
These authors predict that the photosphere of the fireball should
radiate a blackbody-like spectrum with a typical temperature of $\sim 10$ keV.
However, they note that the opacity of a fireball in a strong magnetic
field is a function of energy, allowing the observer to see deeper into the 
fireball at lower energies. 
This has the double effect of broadening the spectral
distribution (extending it towards lower energies), and lowering the
peak of the spectrum.
Magnetic photon splitting in the strong magnetic field of magnetars
further complicates the observed spectrum (Baring 1995, Lyubarsky 2002).
This effect, like double Compton scattering, provides a source
of additional photons, allowing a very efficient thermalization of the fireball.
It also modifies the shape of the radiated spectrum.
Given the importance of this process, it is interesting to investigate whether it
can degrade the energy of the photons from a trapped fireball down to a few keV.
This issue has been studied by Baring (1995), and also by
Thompson \& Duncan (1995), who find that it can 
degrade the energy of the photons down to 10 keV, but probably not
below.  This is however a very complex issue, involving both the details of
the radiation processes and the geometry of the emitting region, and it
deserves additional investigation in view of the present observations.

\begin{figure}
\epsscale{0.6}
\plotone{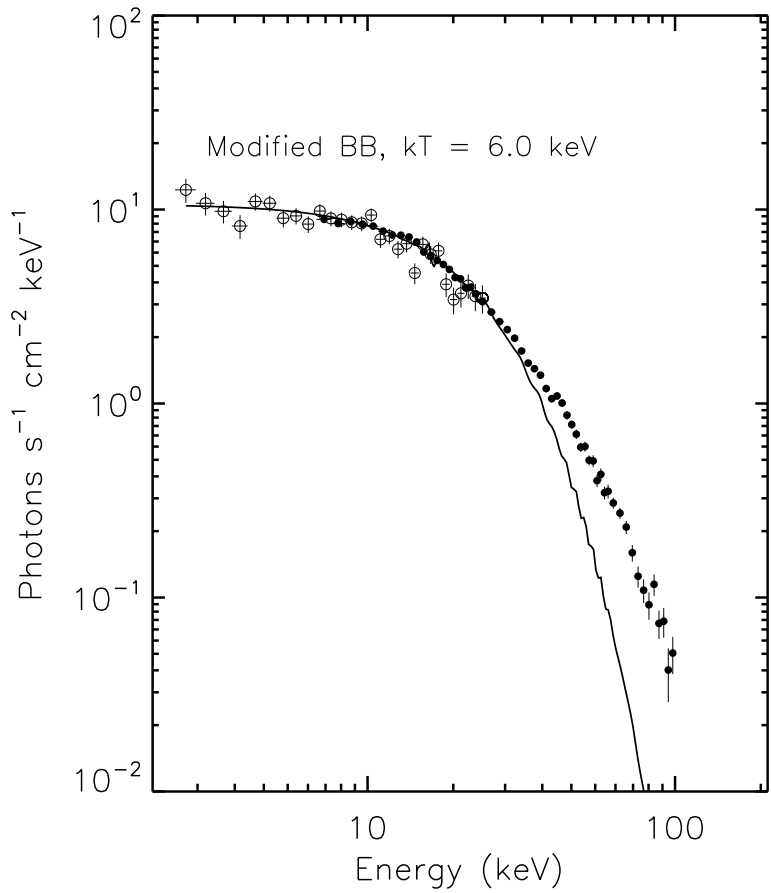}
\caption{Comparison between the modified blackbody spectrum proposed by
Lyubarsky (2002), with an effective temperature of 6 keV, and the observed spectrum.  
Although photon splitting has been taken into account, the observed spectrum 
only fits the modified spectrum below 30 keV.
\label{f8}}
\end{figure}

\begin{figure}
\epsscale{1.1}
\plotone{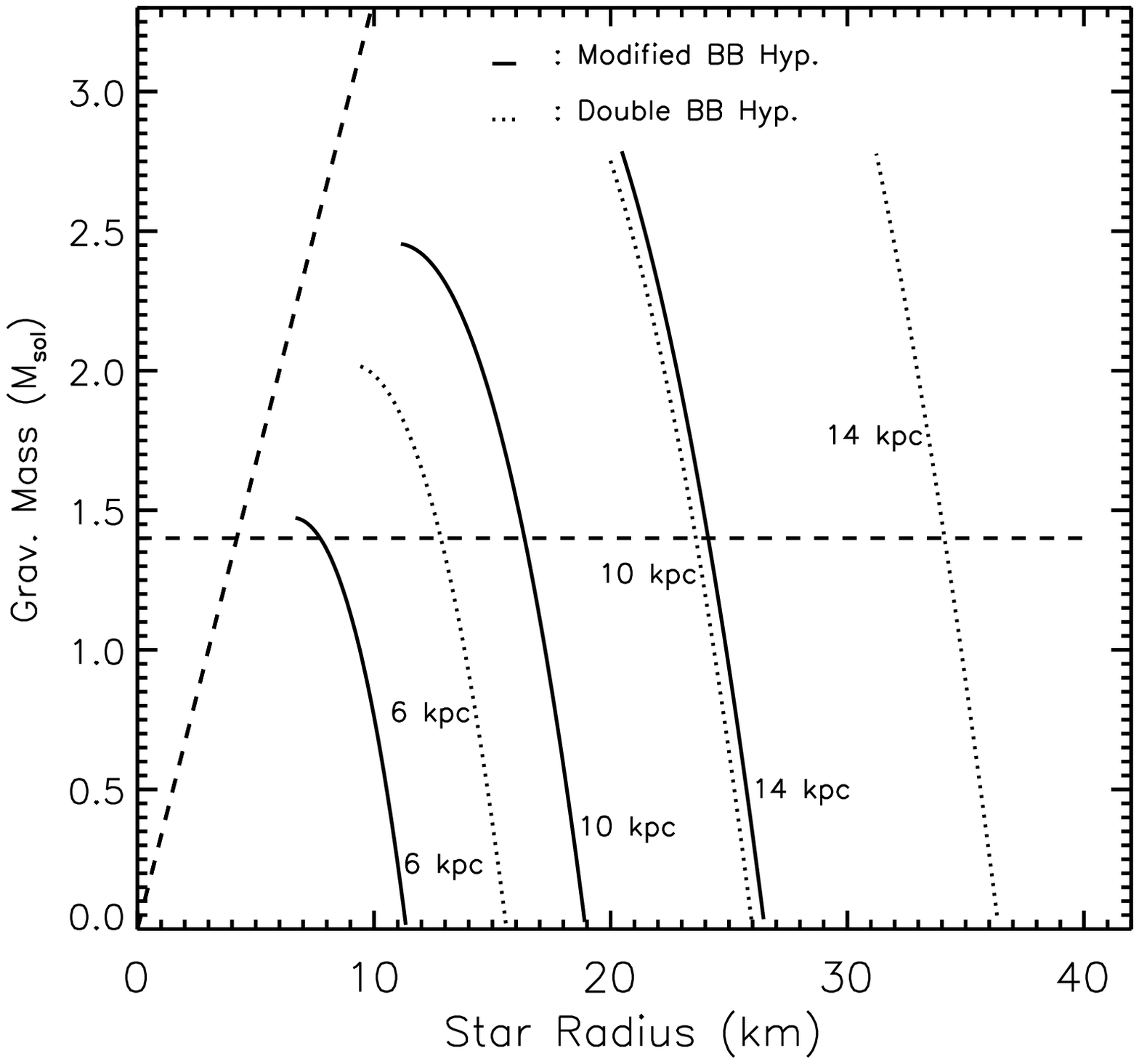}
\caption{Magnetar mass-radius relation for various distances and models.
The dotted line is for the two blackbody (LTBB + HTBB) hypothesis.  The
solid line is for the modified blackbody spectrum of Lybarsky (2002) with
an effective temperature of 6 keV.
The horizontal dashed line indicates a mass of  1.4 \msol.
There are no solutions in the region to the left of the other dashed line.
Solutions exist for source distances of $\sim 6-10$ kpc with inferred radii which
are compatible with 
the accepted values of neutron star radii ($\sim 12-18$ km) 
\label{f9}}
\end{figure}

Although the present event is not a short burst, we checked the agreement 
of our data with these theoretical predictions
by comparing the WXM+FREGATE spectrum with the spectrum predicted
by Lyubarsky. Figure \ref{f8} shows a good agreement between 
the observations and the theory below 30 keV.  Above 30 keV, the model
under-predicts the observed counts, even though the effects of photon splitting
have been included.  The addition of a 10 keV blackbody as a second component 
would rectify this (not shown).
We can use the spectral fit below 30 keV
to infer the effective temperature T$_{\mathrm eff}$ of the fireball. 
We find an effective temperature of 6 keV.
This value allows us to compute the size of the emitting region,
which is $A = L / \sigma T_{\mathrm eff}^{4}$, assuming isotropic emission.
Taking   L = 6 10$^{40}$ erg cm$^{-2}$ s$^{-1}$, and T$_{\mathrm eff}$ = 6 keV,
gives A = 4500 km$^2$, which corresponds to the surface of a sphere of radius 18.9 km,
assuming a distance of 10 kpc, and ignoring the effect of the gravitational
redshift.

The fact that this value is close to the radii of known neutron stars
further reinforces this interpretation. Figure \ref{f9} shows
the mass-radius relation of the magnetar in this interpretation.
We finally note that the observed luminosity of the burst (L = 6 10$^{40}$ erg cm$^{-2}$ s$^{-1}$)
agrees well with the {\it magnetic Eddington luminosity}
discussed by Ulmer (1994) and by Thompson \& Duncan (1995, Section 3.1).
Despite the good agreement with theoretical predictions,
we believe that it is essential to study more bursts, emitted by
several sources, to definitely validate this interpretation of the data.



\section{Conclusions}

We have presented HETE-FREGATE and WXM observations of an intermediate burst
from SGR1900+14.  This is the first time that time-resolved spectral
analysis of such a burst has been possible over a wide energy range,
with good spectral and temporal resolution.
We have found that the spectrum is well described by the sum of two blackbodies. 
One possible interpretation is that we are observing two distinct emitting 
volumes.  Another is that we are observing emission which is 
affected by radiation transfer effects in a superstrong magnetic field.
Both interpretations are consistent with the basic features of the
magnetar model of soft gamma repeaters.  However, it is not possible
to choose between the two on the basis of a single observation.  Fortunately,
HETE-2 has observed numerous SGR bursts from both SGR1900+14 and SGR1806-20.
Analysis of these observations is underway, and may provide less ambiguous
support for one of these hypotheses.

\acknowledgments

We are grateful to R. Perna for extending her magnetar spectrum model
to the FREGATE energy range and making the results available to us,
to C. Guidorzi for giving us detailed information about the BeppoSAX
data, and to M. Baring for useful comments.
KH is grateful for HETE support under MIT contract MIT-SC-R-293291.

\clearpage

\clearpage

\clearpage

\clearpage

\clearpage

\clearpage

\clearpage

\clearpage

\clearpage

\clearpage 

\begin{deluxetable}{cccccccccc}
\tabletypesize{\footnotesize}
\tablecaption{Parameters of the two-blackbody fits. Intervals I1 to I9 correspond 
to those of Figure \ref{f1}. The columns contain the name of the interval,
the duration of the interval, the LTBB temperature, radius and bolometric luminosity,
the HTBB temperature, radius and bolometric luminosity, the number of degrees of freedom (dof)
of the fit, and the Chi square per dof.
The subscript LTBB (HTBB) corresponds to the Low (High) Temperature Blackbody component.
The error bars are for the 90\% confidence level.}
\tablewidth{0pt}
\tablehead{
\colhead{Interval}&\colhead{Duration}&\colhead{kT$_{LTBB}$}&\colhead{R$_{LTBB}$}&\colhead{L$_{LTBB}$}&\colhead{kT$_{HTBB}$}&\colhead{R$_{HTBB}$}&\colhead{L$_{HTBB}$}&\colhead{dof}&\colhead{$\chi ^2$/dof} \\
\colhead{}&\colhead{sec}&\colhead{keV}&\colhead{km}&\colhead{10$^{40}$ erg s$^{-1}$}&\colhead{keV}&\colhead{km}&\colhead{10$^{40}$ erg s$^{-1}$}&\colhead{}&\colhead{} \\
}
\startdata

   I1-I8\tablenotemark{a}  & 4.093   & $ 4.25^{+  0.1}_{-  0.1}$ & $26.30^{+  0.9}_{-  0.8}$ & $ 2.92 \pm   0.09$ & $ 9.78^{+  0.3}_{-  0.3}$ & $ 4.69^{+  0.4}_{-  0.4}$ & $ 2.59 \pm   0.09$ & 99  & $  1.12\tablenotemark{c}$ \\
   I1-I8\tablenotemark{b}  & 4.093 & $ 4.31^{+  0.1}_{-  0.1}$ & $25.83^{+  0.7}_{-  0.7}$ & $ 3.00 \pm   0.09$ & $ 9.88^{+  0.3}_{-  0.3}$ & $ 4.54^{+  0.4}_{-  0.3}$ & $ 2.56 \pm   0.09$ & 66 & $  0.83\tablenotemark{c}$ \\
   I1\tablenotemark{b}  & 0.045 & $ 2.85^{+  1.1}_{-  0.8}$ & $51.00^{+ 29.4}_{- 18.4}$ & $ 2.22 \pm   0.56$ & $ 7.73^{+  1.6}_{-  0.9}$ & $10.39^{+  2.7}_{-  4.3}$ & $ 5.01 \pm   0.71$ & 66 & $  1.03$ \\
   I2\tablenotemark{b}  & 0.030 & $ 6.11^{+  0.4}_{-  0.6}$ & $28.17^{+  2.8}_{-  2.3}$ & $14.38 \pm   1.59$ & $14.01^{+  6.9}_{-  4.0}$ & $ 2.61^{+  4.3}_{-  1.7}$ & $ 3.42 \pm   1.53$ & 66 & $  0.50$ \\
   I3\tablenotemark{b}  & 0.073 & $ 5.49^{+  0.4}_{-  0.4}$ & $29.17^{+  2.2}_{-  2.0}$ & $10.08 \pm   2.13$ & $11.07^{+  2.2}_{-  1.6}$ & $ 4.58^{+  3.0}_{-  2.0}$ & $ 4.09 \pm   2.17$ & 66 & $  1.23$ \\
   I4\tablenotemark{b}  & 0.891 & $ 4.45^{+  0.2}_{-  0.2}$ & $28.42^{+  1.5}_{-  1.3}$ & $ 4.13 \pm   0.29$ & $ 9.27^{+  0.4}_{-  0.4}$ & $ 7.07^{+  0.9}_{-  0.9}$ & $ 4.79 \pm   0.31$ & 66 & $  1.15$ \\
   I5\tablenotemark{b}  & 0.892 & $ 4.29^{+  0.2}_{-  0.2}$ & $26.89^{+  1.4}_{-  1.3}$ & $ 3.18 \pm   0.17$ & $ 9.70^{+  0.6}_{-  0.5}$ & $ 4.42^{+  0.8}_{-  0.7}$ & $ 2.25 \pm   0.18$ & 66 & $  0.94$ \\
   I6\tablenotemark{b}  & 0.891 & $ 4.05^{+  0.2}_{-  0.2}$ & $27.02^{+  1.6}_{-  1.4}$ & $ 2.56 \pm   0.12$ & $ 9.87^{+  0.6}_{-  0.6}$ & $ 3.88^{+  0.7}_{-  0.6}$ & $ 1.86 \pm   0.13$ & 66 & $  0.99$ \\
   I7\tablenotemark{b}  & 0.892 & $ 3.86^{+  0.2}_{-  0.2}$ & $26.36^{+  2.0}_{-  1.7}$ & $ 2.01 \pm   0.09$ & $10.17^{+  0.6}_{-  0.6}$ & $ 3.52^{+  0.6}_{-  0.5}$ & $ 1.72 \pm   0.10$ & 66 & $  0.82$ \\
   I8\tablenotemark{b}  & 0.379 & $ 3.68^{+  0.5}_{-  0.6}$ & $17.77^{+  4.4}_{-  2.8}$ & $ 0.76 \pm   0.21$ & $ 7.71^{+  1.4}_{-  1.1}$ & $ 3.92^{+  2.1}_{-  1.6}$ & $ 0.70 \pm   0.23$ & 66 & $  1.22$ \\
   I9\tablenotemark{b}  & 1.94  & -                         &         -                 &     -              & $ 6.23^{+  0.6}_{-  0.6}$ & $ 3.12^{+  0.7}_{-  0.5}$ & $ 0.19 \pm   0.03$ & 68 & $  0.72$ \\
        
\enddata
\tablenotetext{a}{Joint spectrum (WXM and FREGATE) in the energy range 2-150 keV. See Figure \ref{f2}}
\tablenotetext{b}{FREGATE-only spectrum in the energy range 7-150 keV.}
\tablenotetext{c}{This spectrum includes  3\% systematic errors.}
\label{tab1}
\end{deluxetable}

\clearpage

\begin{deluxetable}{cccccccccc}
\tabletypesize{\footnotesize}
\tablecaption{Parameters of the two-blackbody fits for the 24 sub-intervals i1 to i24 (see text).
The columns contain the name of the interval,
the duration of the interval, the LTBB temperature, radius and bolometric luminosity,
the HTBB temperature, radius and bolometric luminosity, the number of degrees of freedom (dof)
of the fit, and the Chi square per dof.
The subscript LTBB (HTBB) corresponds to the Low (High) Temperature BlackBody component.
The error bars are for the 90\% confidence level.}
\tablewidth{0pt}
\tablehead{
\colhead{Interval}&\colhead{Duration}&\colhead{kT$_{LTBB}$}&\colhead{R$_{LTBB}$}&\colhead{L$_{LTBB}$}&\colhead{kT$_{HTBB}$}&\colhead{R$_{HTBB}$}&\colhead{L$_{HTBB}$}&\colhead{dof}&\colhead{$\chi ^2$/dof} \\
\colhead{}&\colhead{sec}&\colhead{keV}&\colhead{km}&\colhead{10$^{40}$ erg s$^{-1}$}&\colhead{keV}&\colhead{km}&\colhead{10$^{40}$ erg s$^{-1}$}&\colhead{}&\colhead{} \\
}
\startdata
 
      i1  & $0.045$ & $ 2.85^{+  1.1}_{-  0.8}$ & $51.00^{+ 29.4}_{- 18.4}$ & $ 2.22 \pm   0.56$ & $ 7.73^{+  1.6}_{-  0.9}$ & $10.39^{+  2.7}_{-  4.3}$ & $ 5.01 \pm   0.71$ & $ 66$ & $  1.03$ \\
      i2  & $0.030$ & $ 6.11^{+  0.4}_{-  0.6}$ & $28.17^{+  2.8}_{-  2.3}$ & $14.38 \pm   1.59$ & $14.01^{+  6.9}_{-  4.0}$ & $ 2.61^{+  4.3}_{-  1.7}$ & $ 3.42 \pm   1.53$ & $ 66$ & $  0.50$ \\
      i3  & $0.073$ & $ 5.49^{+  0.4}_{-  0.4}$ & $29.17^{+  2.2}_{-  2.0}$ & $10.08 \pm   2.13$ & $11.07^{+  2.2}_{-  1.6}$ & $ 4.58^{+  3.0}_{-  2.0}$ & $ 4.09 \pm   2.17$ & $ 66$ & $  1.23$ \\
      i4  & $0.091$ & $ 3.99^{+  1.1}_{-  1.1}$ & $31.78^{+ 13.5}_{-  5.4}$ & $ 3.32 \pm   1.24$ & $ 7.89^{+  1.6}_{-  0.7}$ & $12.63^{+  4.0}_{-  5.8}$ & $ 8.04 \pm   1.33$ & $ 66$ & $  0.76$ \\
      i5  & $0.101$ & $ 4.64^{+  0.8}_{-  1.0}$ & $27.48^{+  5.5}_{-  2.9}$ & $ 4.54 \pm   1.91$ & $ 8.22^{+  2.2}_{-  1.0}$ & $ 9.69^{+  5.3}_{-  5.5}$ & $ 5.57 \pm   1.95$ & $ 66$ & $  0.84$ \\
      i6  & $0.111$ & $ 4.12^{+  0.6}_{-  0.6}$ & $30.24^{+  5.9}_{-  4.0}$ & $ 3.42 \pm   0.86$ & $ 8.57^{+  0.8}_{-  0.7}$ & $ 9.32^{+  2.4}_{-  2.3}$ & $ 6.10 \pm   0.90$ & $ 66$ & $  0.74$ \\
      i7  & $0.119$ & $ 4.66^{+  0.5}_{-  0.6}$ & $27.39^{+  3.6}_{-  2.6}$ & $ 4.61 \pm   1.39$ & $ 8.89^{+  1.6}_{-  1.1}$ & $ 6.97^{+  3.5}_{-  2.9}$ & $ 3.95 \pm   1.43$ & $ 66$ & $  0.69$ \\
      i8  & $0.130$ & $ 3.91^{+  0.6}_{-  0.6}$ & $32.28^{+  7.0}_{-  4.5}$ & $ 3.17 \pm   0.66$ & $ 8.39^{+  1.0}_{-  0.8}$ & $ 8.57^{+  2.6}_{-  2.5}$ & $ 4.73 \pm   0.69$ & $ 66$ & $  0.73$ \\
      i9  & $0.131$ & $ 4.16^{+  0.5}_{-  0.5}$ & $29.80^{+  4.9}_{-  3.6}$ & $ 3.45 \pm   0.61$ & $ 9.44^{+  1.0}_{-  0.8}$ & $ 6.75^{+  1.9}_{-  1.7}$ & $ 4.71 \pm   0.67$ & $ 66$ & $  0.89$ \\
     i10  & $0.138$ & $ 4.65^{+  0.7}_{-  1.1}$ & $25.05^{+  7.2}_{-  3.0}$ & $ 3.82 \pm   0.94$ & $ 9.30^{+  2.4}_{-  1.6}$ & $ 6.12^{+  4.8}_{-  3.1}$ & $ 3.64 \pm   0.99$ & $ 66$ & $  1.10$ \\
     i11  & $0.150$ & $ 3.72^{+  0.6}_{-  0.5}$ & $32.42^{+  7.6}_{-  5.2}$ & $ 2.61 \pm   0.43$ & $ 8.81^{+  1.0}_{-  0.8}$ & $ 7.31^{+  2.1}_{-  2.0}$ & $ 4.19 \pm   0.46$ & $ 66$ & $  1.18$ \\
     i12  & $0.158$ & $ 4.86^{+  0.3}_{-  0.4}$ & $24.93^{+  2.4}_{-  1.9}$ & $ 4.49 \pm   0.56$ & $11.81^{+  2.4}_{-  2.1}$ & $ 2.63^{+  1.8}_{-  1.0}$ & $ 1.75 \pm   0.56$ & $ 66$ & $  0.97$ \\
     i13  & $0.184$ & $ 4.07^{+  0.3}_{-  0.4}$ & $29.02^{+  3.6}_{-  2.9}$ & $ 3.01 \pm   0.57$ & $ 8.84^{+  1.0}_{-  0.9}$ & $ 5.51^{+  1.9}_{-  1.5}$ & $ 2.41 \pm   0.59$ & $ 66$ & $  0.80$ \\
     i14  & $0.199$ & $ 4.12^{+  0.5}_{-  0.6}$ & $27.31^{+  4.8}_{-  3.0}$ & $ 2.79 \pm   0.62$ & $ 8.39^{+  1.6}_{-  1.3}$ & $ 5.67^{+  3.6}_{-  2.4}$ & $ 2.07 \pm   0.63$ & $ 66$ & $  0.75$ \\
     i15  & $0.218$ & $ 4.20^{+  0.4}_{-  0.4}$ & $25.31^{+  3.3}_{-  2.5}$ & $ 2.60 \pm   0.50$ & $ 8.80^{+  1.3}_{-  1.1}$ & $ 4.94^{+  2.3}_{-  1.6}$ & $ 1.90 \pm   0.51$ & $ 66$ & $  0.88$ \\
     i16  & $0.219$ & $ 4.22^{+  0.4}_{-  0.5}$ & $25.63^{+  3.4}_{-  2.5}$ & $ 2.72 \pm   0.53$ & $ 8.99^{+  1.7}_{-  1.3}$ & $ 4.53^{+  2.6}_{-  1.8}$ & $ 1.74 \pm   0.54$ & $ 66$ & $  0.72$ \\
     i17  & $0.221$ & $ 4.10^{+  0.3}_{-  0.3}$ & $27.70^{+  3.0}_{-  2.5}$ & $ 2.83 \pm   0.32$ & $10.15^{+  1.5}_{-  1.3}$ & $ 3.37^{+  1.5}_{-  1.1}$ & $ 1.56 \pm   0.33$ & $ 66$ & $  0.92$ \\
     i18  & $0.235$ & $ 4.17^{+  0.3}_{-  0.4}$ & $25.03^{+  3.3}_{-  2.5}$ & $ 2.46 \pm   0.34$ & $ 9.74^{+  1.2}_{-  1.1}$ & $ 3.95^{+  1.5}_{-  1.1}$ & $ 1.83 \pm   0.34$ & $ 66$ & $  0.66$ \\
     i19  & $0.232$ & $ 3.68^{+  0.3}_{-  0.3}$ & $30.24^{+  4.3}_{-  3.5}$ & $ 2.17 \pm   0.29$ & $ 9.01^{+  1.0}_{-  0.8}$ & $ 4.98^{+  1.4}_{-  1.2}$ & $ 2.12 \pm   0.32$ & $ 66$ & $  0.91$ \\
     i20  & $0.259$ & $ 3.50^{+  0.4}_{-  0.4}$ & $30.32^{+  5.4}_{-  4.0}$ & $ 1.79 \pm   0.24$ & $ 8.42^{+  0.9}_{-  0.7}$ & $ 5.56^{+  1.6}_{-  1.4}$ & $ 2.02 \pm   0.26$ & $ 66$ & $  1.23$ \\
     i21  & $0.271$ & $ 3.69^{+  0.3}_{-  0.3}$ & $27.44^{+  4.4}_{-  3.4}$ & $ 1.81 \pm   0.18$ & $ 9.86^{+  0.9}_{-  0.8}$ & $ 4.05^{+  1.0}_{-  0.8}$ & $ 2.02 \pm   0.20$ & $ 66$ & $  0.80$ \\
     i22  & $0.394$ & $ 4.14^{+  0.2}_{-  0.3}$ & $23.92^{+  2.5}_{-  2.0}$ & $ 2.18 \pm   0.16$ & $10.94^{+  1.5}_{-  1.3}$ & $ 2.56^{+  1.0}_{-  0.7}$ & $ 1.22 \pm   0.17$ & $ 66$ & $  0.84$ \\
     i23  & $0.379$ & $ 3.68^{+  0.5}_{-  0.6}$ & $17.77^{+  4.4}_{-  2.8}$ & $ 0.76 \pm   0.21$ & $ 7.71^{+  1.4}_{-  1.1}$ & $ 3.92^{+  2.1}_{-  1.6}$ & $ 0.70 \pm   0.23$ & $ 66$ & $  1.22$ \\
     i24  & $1.940$ & -                         &         -                 &     -              & $ 6.23^{+  0.6}_{-  0.6}$ & $ 3.12^{+  0.7}_{-  0.5}$ & $ 0.19 \pm   0.03$ & 68 & $  0.72$ \\

\enddata
\label{tab2}
\end{deluxetable}



\begin{thebibliography}{}

\bibitem{apte01} Aptekar, R.L., Frederiks, D.D., Golenetskii, S.V. et al. 2001, \apjs, 137, 227

\bibitem{arna96} Arnaud, K.A. 1996, in Astronomical Data Analysis Software and Systems V, 
           eds. Jacoby G. and Barnes J., p. 17, ASP Conf. Series volume 101

\bibitem{atte03} Atteia, J.-L., Boer, M., Cotin, F. et al. 2003, 
           in Gamma-Ray Burst and Afterglow Astronomy 2001, Eds.
           G. Ricker and R. Vanderspek, AIP Conference Proceedings 662 (AIP: New York), 17


\bibitem{bari95} Baring, M., et al. 1995, \apj \, 440, L69

\bibitem{dunc92} Duncan, R., \& Thompson, C. 1992, \apj \, 392, L9

\bibitem{feni94} Fenimore, E., Laros, J., and Ulmer, A. 1994, \apj \, 432, 742

\bibitem{fero99} Feroci, M., Frontera, F., Costa, E., Amati, L., Tavani, M., Rapisarda, M., and Orlandini, M.
           1999, \apj \, 515, L9

\bibitem{fero01a} Feroci, M., in't Zand, J., Soffitta, P., Hurley, K., Frontera, F., and Mazets, E. 
           2001, GCN Circ. 1060 (http://gcn.gsfc.nasa.gov/gcn/gcn3/1060.gcn3)

\bibitem{fero01b} Feroci, M., Hurley, K., Duncan, R., and Thompson, C. 2001, \apj \, 549, 1021

\bibitem{fero03} Feroci, M. et al., 2003, \apj, 596, 470

\bibitem{fero04} Feroci, M. Caliandro, G.A., Massaro, E., Mereghetti, S., \& Woods, P. 2004, \apjl ,
in press, (astro-ph/0405104) 

\bibitem{frai99} Frail, D., Kulkarni, S., and Bloom, J. 1999,  \nat \, 398, 127

\bibitem{gole84} Golenetskii, S., Ilyinskii, V., and Mazets, E. 1984 \nat \, 307, 41

\bibitem{guid01} Guidorzi, C. et al. 2001, GCN Circ. 1041 (http://gcn.gsfc.nasa.gov/gcn/gcn3/1041.gcn3)

\bibitem{guid03} Guidorzi, C., Frontera, F., Montanari, E., Feroci, M., Amati, L., 
Costa, E., and Orlandini, M. 2004, A\&A 416, 293

\bibitem{hurl94} Hurley, K., et al. 1994, \apj \, 431, L31

\bibitem{hurl96} Hurley, K., et al. 1996, \apj \, 463, L13

\bibitem{hurl99a} Hurley, K., et al. 1999a, \apj \, 510, L107

\bibitem{hurl99b} Hurley, K., et al. 1999b, \apj \, 510, L111

\bibitem{hurl99c} Hurley, K., et al. 1999c, \nat \, 397, 41

\bibitem{hurl01a} Hurley, K., Montanari, E., Guidorzi, C., Frontera, F., and Feroci, M. 2001a,
           GCN Circ. 1043 (http://gcn.gsfc.nasa.gov/gcn/gcn3/1043.gcn3)

\bibitem{hurl01b} Hurley, K., Cline, T., Mazets, E., and Golenetskii, S. 2001b, GCN Circ. 1071
           (http://gcn.gsfc.nasa.gov/gcn/gcn3/1071.gcn3)

\bibitem{ibra01} Ibrahim, A., Strohmayer, T., Woods, P., et al. 2001, \apj \, 558, 237

\bibitem{kouv93} Kouveliotou, C., et al. 1993, \nat \, 362, 728

\bibitem{kouv99} Kouveliotou, C., et al. 1999, \apj \, 510, L115

\bibitem{kouv01} Kouveliotou, C., et al. 2001, \apj \, 558, L47

\bibitem{lyu02} Lyubarsky, Y. 2002, \mnras \,332, 199

\bibitem{maze79} Mazets, E., Golenetskii, S., and Guryan, Yu. 1979, 
           Sov. Astron. Lett., 5(6), 343

\bibitem{maze99} Mazets, E., et al. 1999, Astron. Lett. 25(10), 635

\bibitem{mont01} Montanari, E., Guidorzi, C., Frontera, F., Calura, F., and Feroci, M. 2001,
           GCN Circ. 1081 (http://gcn.gsfc.nasa.gov/gcn/gcn3/1081.gcn3)

\bibitem{oliv03} Olive, J.-F. et al. 2003, in in Gamma-Ray Burst and Afterglow Astronomy 2001, Eds.
           G. Ricker and R. Vanderspek, AIP Conference Proceedings 662 (AIP: New York), 88

\bibitem{pacz92} Paczy\'nski, B. 1992, Acta Astronomica 42, 145

\bibitem{pern01} Perna, R., Heyl, J., Hernquist, L., Juett, A., and Chakrabarty, D. 2001, 
           \apj \, 557, 18

\bibitem{rick01a} Ricker, G., et al. 2001a, GCN Circ. 1073 (http://gcn.gsfc.nasa.gov/gcn/gcn3/1073.gcn3)

\bibitem{rick01b} Ricker, G., et al. 2001b, GCN Circ. 1074 (http://gcn.gsfc.nasa.gov/gcn/gcn3/1074.gcn3)

\bibitem{rick01c} Ricker, G., et al. 2001c, GCN Circ. 1078 (http://gcn.gsfc.nasa.gov/gcn/gcn3/1078.gcn3)

\bibitem{thom95} Thompson, C., \& Duncan, R. 1995, \mnras \, 275, 255

\bibitem{thom96} Thompson, C., \& Duncan, R. 1996, \apj \, 473, 322

\bibitem{thom01} Thompson, C., \& Duncan, R. 2001, \apj \, 561, 980

\bibitem{tori03} Torii, K., Yoshida, A., Kawai, N. et al. 2003, 
           in Gamma-Ray Burst and Afterglow Astronomy 2001, Eds.
           G. Ricker and R. Vanderspek, AIP Conference Proceedings 662 (AIP: New York), 111

\bibitem{ulm94} Ulmer, A. 1994, \apj \, 437, L111

\bibitem{vasi94} Vasisht, G., Kulkarni, S., Frail, D., and Greiner, J. 1994, 
           \apj \, 431, L35

\bibitem{vrba00} Vrba, F., Henden, A., Luginbuhl, C., Guetter, H., Hartmann, D., and Klose, S. 2000, 
           \apj \, 533, L17

\bibitem{wood01} Woods, P., Kouveliotou, C., and Gogus, E. 2001, GCN Circ. 1056 
           (http://gcn.gsfc.nasa.gov/gcn/gcn3/1056.gcn3)

\bibitem{wood03} Woods, P., et al. 2003, \apj \, 596, 464

\end{thebibliography}
\end{document}